\documentclass[a4paper,11pt,twocolumn]{quantumarticle}
\pdfoutput=1
\usepackage{bm}
\usepackage{amsmath}
\usepackage{amssymb}
\usepackage{graphicx}
\usepackage{IEEEtrantools}
\usepackage{epsfig}
\usepackage{epstopdf}

\usepackage{hyperref}

\begin{document}

\newtheorem{theorem}{Theorem}
\newtheorem{corrolary}{Corollary}

\def\pr{\prime}
\def\be{\begin{equation}}
\def\en#1{\label{#1}\end{equation}}
\def\dag{\dagger}
\def\bar#1{\overline #1}
\def\U{\mathcal{U}}
\newcommand{\per}{\mathrm{per}}
\newcommand{\rd}{\mathrm{d}}
\newcommand{\vare}{\varepsilon }

 \newcommand{\m}{\mathbf{m}}
\newcommand{\n}{\mathbf{n}}
\newcommand{\s}{\mathbf{s}}
\newcommand{\bk}{\mathbf{k}}
\newcommand{\bl}{\mathbf{l}}
\newcommand{\br}{\mathbf{r}}

\newcommand{\Lim}[1]{\raisebox{0.5ex}{\scalebox{0.8}{$\displaystyle \lim_{#1}\;$}}}

\title{  \href{https://doi.org/10.22331/q-2021-03-29-423}{Boson sampling  cannot be faithfully simulated by   only the   lower-order multi-boson interferences }}

  \author{Valery  Shchesnovich } 
  \email{valery069@gmail.com}
 \begin{abstract} 
To simulate noisy boson sampling  approximating it by only the   lower-order multi-boson interferences (e.g., by a smaller number of interfering bosons and classical particles) is very  popular idea.  I show that  the output data from  any such  classical simulations    can be efficiently distinguished from that of the quantum device they try to simulate, even with  finite  noise in the latter.         The  distinguishing datasets    can be the experimental  estimates of   some large    probabilities,  a wide     class of such is  presented.    This   is a sequel    of  \textit{Quantum} \textbf{5},   423 (2021), where I   present  more accessible    account of the  main result  enhanced by  additional insight on the contribution from the  higher-order multi-boson interferences in presence of noise. \end{abstract}
\maketitle

 \section{Introduction}

Presence of   weak  noise is accounted in the boson sampling  idea \cite{AA}. For $N$ interfering bosons  the  number of  classical computations    required for  the  noiseless  boson sampling is  estimated to be   $O(N2^N)$ \cite{Cliffords}. However, a polynomial in $N$  classical simulation   could become   possible for some strong enough    noise.  In the experiments  \cite{20ph60mod,ExpGBS2,GBS144}   there is  some amount of noise. Is it  weak noise  or strong noise?    

Due to  an exponentially large in $N$  space of outcomes (exponentially small probabilities) the experimentalists check only some lower-order correlations.   The universal feature of noise effect, be it photon  distinguishability \cite{VS14}, photon  losses \cite{Brod},  or  unstable noisy  network   \cite{LP,KK,Arkh}, is that    the   higher  orders  of  multi-boson interference (precise definition is given below)  affected stronger by   presence of    noise  (see also  the discussion  \cite{Bblog}).    Efficient classical approximations    exploit  this   effect of  noise   \cite{K1,R1,OB,PRS,NonUnifLoss,RSP,VS2019,MRRT}.   When   amplitudes  of noise  in boson sampling are not scaling down    with its size, one can call such noise finite nose. It is known that for such  noise the  correlation between  the output  distribution of noiseless  boson sampling and    that of the  noisy one   tend to zero  \cite{KK} and   the total variation distance between the two distributions   cannot be small  \cite{VS2019}.  Efficient classical approximation of boson sampling   can be constructed for finite noise amplitudes by   imposing a cutoff at a fixed  order $K=O(1)$ (as $N$ scales up) of multi-boson interferences  \cite{KK,R1,RSP,VS2019,MRRT}  (see also   explanation in Aaronson's blog Ref. \cite{Ablog}).  Such lower-order approximations are the focus of this work. 

  I show  that    classical simulations   accounting for only the  lower-order  multi-boson interferences   can be efficiently distinguished from the quantum  device with finite noise  they try to simulate.         Despite the   fact that the sensitivity to noise  is  proportional to the order of  quantum correlations,   the higher-order correlations (higher-order multi-boson interferences)  still make up the difference. Thus experimentalists should find ways to check the higher-order correlations. One way is presented below.    This is   accessible   exposition  of    the main result in  the recent paper  \cite{myQuantum}  enhanced by   additional insight on the contribution from the higher orders of    multi-boson interferences in the presence of noise.  

The text is structured as follows.  I  present and discuss  the main result of Ref. \cite{myQuantum} (section  \ref{sec2}), then give an additional insight on the higher-order correlations in the presence of noise (section \ref{sec3}), and then discuss the relation to classical simulations and the recent experiment on boson sampling (section \ref{sec4}). Conclusion (section \ref{sec5}) contains the main message to be taken from this  text. 

 \section{How to distinguish noisy boson sampling from classical approximations}
\label{sec2}

The    computational complexity of boson sampling is related to the fact that the  number of possible ``quantum paths" in  the quantum transition of $N$ bosons through a unitary linear interferometer is exponential in $N$, due to bosons being identical particles.  Different  quantum paths of $N$ bosons are given by different permutations of bosons  by the symmetric group of $N$ objects,  each  particular path of bosons is composed of multi-boson interferences. Let us  define what is meant by  the ``multi-boson  interferences of order  $\ell$".  To this goal, we   partition  the full multi-boson interference  into  some disjoint  classes.    As is known, a permutation  can be decomposed into a product of disjoint cycles. A permutation  $\pi\in S_N$ of $X\equiv (1,\ldots,N)$ is a cycle of length $\ell$  if it \textit{cyclically} permutes  $\ell$ elements, e.g.,  $i_1\to i_2\to \ldots \to i_\ell\to i_1$ ($  i_\alpha \in X$, $i_\alpha\ne i_\beta$ for $\alpha\ne \beta$).  The key  property of the disjoint  cycles is  that  the  cycles    of length $\ell$    map to  the  multi-boson interferences of exactly $\ell$ bosons  \cite{Ninter}. For example, cycles of length $\ell=1$,  i.e., the fixed points,   correspond to  ``lone", or classical, bosons, not interfering with other bosons,  and cycles of length $\ell=2$ correspond to two-boson   interferences, responsible for the second-order correlations.  Moreover,  the quantum multi-boson correlation function  of    order $K$,  which describes the  joint detection of only  $K$ bosons out of $N$ (averaging on $N-K$ bosons in the output probability),   depend only on the cycles  (the multi-boson interferences) of orders $\ell\le  K$ \cite{Ninter}. 

Bosons are indistinguishable, i.e., can interfere on a unitary interferometer,   only to  the degree given by the   overlap of  their states  in the  internal degrees of freedom, called the internal states   (for photons there are infinitely many of such, due to the  continuous  spectral shape).    Assuming a constant overlap $0<\xi\le 1$  of the internal states of two bosons, one can show that multi-boson interferences of orders $\ell \ge 2$ (given by the cycles of length $\ell$) acquire a weight equal to $\xi^\ell$   \cite{myQuantum}  (for $\ell=1$, i.e., for the lone bosons,   the weight is equal to $1$; the output probability formula is  reproduced in appendix \ref{appB}).     

Let us define   ``the lower-order  interferences"  by combining all the  orders $\ell$  of   multi-boson interferences satisfying $\ell\le K$, with some fixed $K = O(1)$   for the total number of bosons  $N$    scaling  to infinity.    Due  to the fact that the higher-order interferences   have  much stronger noise sensitivity then the lower-order ones,     it   seems   reasonable enough  to assume that   the former  do not  matter in the presence of finite    noise and, therefore, one could faithfully approximate  such noisy   boson sampling  device  by accounting only for the latter.    Will this simple idea work?   Below I argue that no, it will not: One    can efficiently distinguish the output data  set coming from such classical simulators  and that from  the boson sampling they try to simulate.    

The above  negative answer applies not only to the   noise  due to partial distinguishability of bosons,  but   also to    other sources of noise and to their combined effect as well. Here we consider also   imperfect transmission (losses) of bosons through the device,  accounted for by a uniform  transmission coefficient $0<\eta\le 1$. The probability of a single boson being transmitted by the  interferometer is $\eta^2$.  The  so-called dark counts of  detectors (not related to  particle detection), which follow   the usual  Poisson distribution     $\pi(n) = \frac{\nu^n}{n!}e^{-\nu}$,  are  also accounted  with some  uniform   rate $\nu$ for all detectors. Moreover, there are  equivalence relations between  action of  various sources of noise on boson sampling (the proof can be found in  Ref. \cite{VS2019}).  For instance,  noise in interferometer  \cite{KK} is equivalent to a combined action of  boson losses exactly compensated by dark counts of detectors (a special case of the shuffled bosons model of Ref.  \cite{Brod}).       We consider  a  linear  interferometer  with a unitary matrix $U_{kl}$, where there are   $M$ input  and output  ports ($1\le k,l\le M$),  and impose no relation between $M$ and $N$,  except that $M\ge N$ (as is common in recent experiments with large numbers of bosons \cite{20ph60mod,ExpGBS2,GBS144}).   Introduce   the density of bosons parameter $\rho = N/M$. 
 
 Our goal is to bound  \textit{from below} the total variational distance between the probability distributions of a noisy boson sampling  $p_\m$  with some noise parameters $\xi, \eta$, $\nu$ (see an example for $\nu=\eta=1$ and arbitrary $\xi$  in appendix \ref{appB}) and a classical simulation  accounting for the lower-order multi-boson interferences  $p^{(K)}_\m$, where such a simulation is obtained by imposing a cutoff at  an order $K=O(1)$  of the multi-boson correlations. One  can impose the cutoff,   for example,   by allowing only  $K$ (random) bosons  to interfere  supplemented by  $N-K$ ``lone" bosons  passing through an interferometer  one by one \cite{R1,RSP,VS2019}. The general way is to  limit the disjoint cycle lengths by $K$  in the symmetric group  $S_N$ describing the multi-boson interferences \cite{myQuantum}.  The total variational distance reads
\be
 \mathcal{D}(p,p^{(K)})= \frac12 \sum_{\m} |p_\m - p^{(K)}_\m|, 
\en{E2}
where the sum runs over all possible  configurations $\m=(m_1,\ldots,m_M)$ of  bosons  in the output ports, with $m_l$ being the number of bosons in output port $l$.  The difference in   probability  of any subset $ \Omega$ of the output configurations $\m$ bounds the total variation distance from below,
\be
\mathcal{D}(p,p^{(K)})\ge \left|  {P}_\Omega-P^{(K)}_\Omega \right|, \quad P_\Omega \equiv \sum_{\m\in \Omega} p_\m.
\en{E3}
Observe that the  equality   is necessarily achieved for a certain subset 
$\Omega_*$  depending  on $U$ and other parameters of the setup. 

Consider first one specific choice of   $\Omega$: the probability to detect  zero bosons at a single output port, say port $l=1$,  and arbitrary numbers of bosons in other output ports,   $m_1=0$, $m_2+\ldots+ m_M=N$.  Denote this probability  by $P_1$ ($P^{(K)}_1$) and the difference  $\Delta P_1\equiv P_1-P^{(K)}_1$. Assuming that $K  \ll \sqrt{N}$ we obtain \cite{myQuantum}:  
\begin{align}
\label{E4}
    & \mathcal{D}(p,p^{(K)})\ge |\langle \Delta P_1\rangle | \approx   \frac{(\xi \eta\rho)^{K+1}}{1+\xi \eta \rho}e^{-1-\nu-\eta \rho}\equiv W_1,  \\
   &\frac{\langle (\Delta P_1)^2 \rangle  - \langle \Delta P_1\rangle^2}{W^2_1} \approx  \frac{(1-\rho)(K+1)^2}{N}. \nonumber 
\end{align}
where the averaging is performed over the Haar-random interferometers $U$\footnote{The approximation sign  in the result in Eq. (\ref{E4})  is a technical issue due to  approximation of a discrete function in natural numbers $K\ge 1$ by a smooth law, the latter      approaches the  original function  for    $K\gg 1$, see Ref. \cite{myQuantum}.}. Observe that, in contrast,      single output probability, i.e., $ p_\m$ and $p^{(K)}_\m$,  vanishes   exponentially  with $N$. 

The vanishing of the  relative variance as $N$ scales up,    given that  $K \ll \sqrt{N}$,   implies (by   the standard  Chebyshev's  inequality)   that the lower bound in Eq. (\ref{E4})   applies almost surely over the  Haar-random interferometers  (only some  subset of  the   interferometers having  vanishing  Haar measure   does  not satisfy the bound).

Furthermore, for a wide  class of interferometers having   one balanced output port $|U_{k,1}| = \frac{1}{\sqrt{M}}$   the difference in probability of no boson counts   in the balanced output port  satisfies \cite{myQuantum}  
\be
\mathcal{D}(p,p^{(K)})\ge |\Delta P_1|\ge W_1\left(1- \left|O\left(\frac{K^2}{N}\right)\right|\right),
\en{E5}
 where the  minus sign indicates a negative correction. Such interferometers contain  a wide class: $U= \mathcal{F}(1\oplus  V)$ with the Fourier interferometer  $\mathcal{F}_{kl} =\frac{1}{\sqrt{M}}e^{2i\pi\frac{kl}{M}}$ and an arbitrary  $(M-1)$-dimensional unitary interferometer $V$.

 If the   parameters    $\xi, \eta,\nu$  remain fixed   when  the total number of bosons $N$ scales up,   then such   boson sampling has finite noise.    In this case the  correlation between  the output  distribution of noiseless  boson sampling and    that of the  noisy one   tend to zero  \cite{KK}.  Moreover,  the total variation distance between the two distributions   cannot be small  \cite{VS2019}.    Finite noise is   experimentally relevant, since  most of the   experimental noise amplitudes remain finite with scaling up the size of a quantum device (except for   the transmission $\eta$    \cite{PRS}).      Eqs. (\ref{E4})-(\ref{E5})  point that   with only a polynomial in $N$ number of runs of  the   quantum  device with  finite noise,  one would have accumulated  a data set  sufficient to distinguish  the output   probability distribution  from   that produced by any  classical simulation  accounting for only the lower-order multi-boson interferences.  Consider  the   no-collision boson sampling with $M \sim  N^2$, i.e. with the  density of bosons $ \rho \sim 1/N$.  In this case the lower bound $W_1$ in Eq. (\ref{E4}) scales as $O(\rho^{K+1}) = O(N^{-K-1})$ in the total number of bosons.  To tell apart the two distributions one has to only estimate   the probability $P_1$  using   $\mathcal{T}\gg N^{K+2}$  output datasets,  to reduce   the statistical error $\mathcal{R}\sim \frac{1}{\sqrt{\mathcal{T}}} $ in them  well below  the lower bound   in Eq. (\ref{E4}).

 In  the  strong collision regime  $M\sim N$ (a finite  density of bosons  $\rho \sim 1$) -- which is  the regime of current boson sampling experiments  \cite{20ph60mod,ClassSimGBS,GBS144} --   we get   $|\langle \Delta P_1\rangle| = O(1)$, i.e.,  the lower bound is independent of the total  number of bosons $N$. In this regime of boson sampling   the approximation    by the lower-order multi-boson interferences would  be exposed   after  a fixed number of runs $\mathcal{T}$ of the quantum device with  arbitrarily  large number of bosons $N$, dependent only on the (finite) noise amplitudes and $K$:  $\mathcal{T} = \mathcal{T}(\xi, \eta,\nu,\rho, K)$.  This fact indicates that the strong collision regime is even worst approximated by taking onto account only the lower-order  multi-boson interferences. 

More generally,    numerical simulations show that a similar  lower bound  as in  Eq. (\ref{E4}) can be  expected  for  the   probability difference  to detect no bosons in  $L<M$  output ports  ($|\Omega| = L$)   \cite{myQuantum}.   Such a probability   can serve     to  rule out  the   classical simulation based on the lower-order interferences.   The   probabilities to detect no bosons in  $L<M$  output ports can be used, therefore,  as witnesses of boson sampling,  they are given by the matrix permanents of  some positive-semi-definite  Hermitian matrices, built from the interferometer matrix, presented in appendix \ref{appC}. Such matrix permanents     can be  efficiently estimated by the algorithm of Ref. \cite{PermHerm}. 
 
 \section{ The  set of lower-order interferences is an  exponentially small  fraction of all multi-boson  interferences    }
 \label{sec3}
 
Additional insight    is provided by the asymptotic estimate on the   fraction  of  the lower-order  multi-boson interferences   in the presence of a finite-amplitude noise,     counting  the  $\ell$th-order interferences with the weight function $\xi^\ell$.  Surprisingly,  even for a finite noise, $\xi^{-1}= O(1)$,   the relative contribution from the lower-order interferences  is vanishing exponentially fast in the total number of bosons $N$. 

Let us first consider  the noiseless case. For  the  total number  $Z^{(K)}_N$ of permutations in $S_N$ decomposable into the disjoint cycles of length $\ell \le K=O(1)$ we get as $N\to \infty$ (see details in appendix \ref{appA})
\begin{eqnarray}
 \label{Nonoise}
&& \mathcal{F}^{(K)}_N\equiv \frac{Z^{(K)}_N}{N!}\\
   &&  < \frac{1+o(1)}{\sqrt{2\pi N}}\exp\left\{ -N\left[\frac{\ln N - 1}{K} -  \frac{e-1}{N^\frac{1}{K}}\right]  \right\}.\nonumber
 \end{eqnarray}
  
In the  presence of noise,  where noise   is  due to   partial distinguishability of bosons  with a uniform overlap $\xi$, we estimate the   ratio of the   number  $\mathcal{Z}^{(K)}_N(\xi)$ of  permutations in $S_N$  decomposable into the  disjoint cycles  of  lengths $\ell \le K$,   weighted by $\xi$ as above,  to all weighted  permutations $\mathcal{Z}_N(\xi)$ (see details in appendix \ref{appAB}):
\begin{eqnarray}
\label{Noise}
&& \!\!\frac{\mathcal{Z}^{(K)}_N(\xi)}{\mathcal{Z}_N(\xi)} < \frac{\mathcal{F}^{(K)}_N}{ \xi^{N}}  \\
 && \!\!< \frac{1+o(1)}{\sqrt{2\pi N}}\exp\left\{ -N\left[\frac{\ln N - 1}{K} -  \frac{e-1}{N^\frac{1}{K}}-\ln \xi^{-1}\right]  \right\}.\nonumber 
 \end{eqnarray}
Therefore,  as  in the  noiseless case   ($\xi = 1$), the fraction    of permutations with the disjoint cycles of lengths not exceeding $K$ is exponentially vanishing in $N$  for $K=O(1)$ and  any finite noise $\xi^{-1} = O(1)$. 

  \section{ Implication for experimental verifications of boson sampling  }
\label{sec4}

The above results   imply   that to  validate   boson sampling  against classical simulations one  must  go beyond the lower-order correlations (involving only the lower-order interferences).    Otherwise an efficient   classical approximation could  be found for the data sets   containing only the  (standard in the field)  lower-order correlations. And indeed,   recently  an  efficient classical simulation was found  \cite{ClassSimGBS} for such  experimental data  of Ref. \cite{GBS144}, namely the marginal probabilities,  depending only     the lower-order interferences (though  the Gaussian  variant of boson sampling was considered, the main idea and the conclusions are expected to hold).  The   above discussion,  however, predicts that  such a   classical simulator  can   be efficiently distinguished from the boson sampling by looking at an output probability of detecting no bosons in a (fixed) subset of output ports, such as in Eq. (\ref{E4}).

 \section{Conclusion}
 \label{sec5}
 
  We have considered the      classical   approximations  of  finite-noise  boson sampling  by the lower-order multi-boson interferences, which includes  the only known to date efficient approximations for a finite noise, e.g., by  a smaller   number  interfering bosons  padded by classical particles. It is argued that the set of  sampling data coming from such approximations  can be efficiently distinguished from the  boson sampling data they  simulate.         The   output probabilities counting no bosons in  fixed subsets of output ports can serve as the efficient distinguishers  between  the output distribution coming from  quantum device and classical simulations.    Surprisingly, for   boson sampling with a large number of bosons   at input on  a smaller-sized  interferometer,  than is required  for the  no-collision regime,  the number of  runs of the quantum device sufficient for distinguishing it from the classical simulations   does not depend on the number of bosons, but  solely  on the amplitudes of noise.   The results  leave an open problem for the future: Can we  classically  simulate, both   efficiently and \textit{faithfully},    the dataset from the boson sampling     with finite   amplitudes of noise, i.e.,  as in the current experiments on boson sampling?
 \section*{Acknowledgements}
   This work   was supported by the National Council for Scientific and Technological Development (CNPq) of Brazil,  Grant 307813/2019-3. 
 
  \appendix
  

  \section{Output probability of  boson sampling  with partially distinguishable bosons   }
  \label{appB}
Consider $N$  single  bosons in internal states $|\psi_1\rangle,\ldots, |\psi_N\rangle$   at inputs $k=1,\ldots, N$ of a  unitary interferometer $U_{kl}$ of size $M$.  Assuming that   $\langle\psi_j|\psi_k\rangle =\xi$, for $j\ne k$,  the   probability $p(\xi) $ to count $m_1,\ldots,m_M$ bosons in the output ports    has the following form (see Ref.  \cite{myQuantum}) 
\begin{eqnarray}
\label{B1}
&& p_\m(\xi) = \frac{1}{\m!} \sum_\sigma\sum_\tau \xi^{N-C_1(\tau\sigma^{-1})} \prod_{i=1}^NU_{\sigma(i),l_i}U^*_{\tau(i),l_i}\nonumber\\
&&  = \frac{1}{\m!} \sum_{\sigma} \sum_\pi \xi^{N-C_1(\pi)} \prod_{i=1}^NU_{\sigma(i),l_i}U^*_{\pi\sigma(i),l_i},
\end{eqnarray}
 where   $\m! = m_1\ldots m_M!$, $1\le l_1\le \ldots\le l_N\le M$ is the multi-set of output ports corresponding to occupations $\m = (m_1,\ldots,m_M)$, $C_1$ is the number of fixed points in permutation, and in the second  expression  we use the  relative permutation $\pi \equiv \tau\sigma^{-1}$.   The weight  $\xi^{N-C_1(\pi)}$  is due to   multi-boson interferences     (the cycles of length $\ell \ge 2$).  

 The expression in Eq. (\ref{B1})   can be understood without any derivation as follows. We have     $N$ input bosons (in an order given by permutation $\sigma$) distributed over at most  $N$  output ports  $1\le l_1\le \ldots \le l_N\le M$ (i.e., some port indices could coincide).   The probability is a sum  over all possible   products  of  the quantum  amplitude,    $ \prod_{i=1}^NU_{\sigma(i),l_i}$,  of the  transition  $\sigma(i)\to l_i$, $i=1,\ldots, N$,  and  the conjugate amplitude  $ \prod_{i=1}^NU^*_{\pi\sigma(i),l_i}$   with arbitrary $\pi$-permuted transition (since  bosons are identical, our label ling them by $\sigma$ has no physical meaning),  weighted by the overlap  of the internal states of bosons $\xi^{N-C_1(\pi)}$ and  divided by the number $\m!$ of identical  terms in the sum  over the permutations $\sigma$ and $\pi$.

 \medskip
 \section{The fraction of permutations   with only lower-order cycles}
 \label{appA}
 
Let us estimate the fraction of permutations, in the  group $S_N$ of permutations of $N$ objects,  that have no disjoint cycle of length greater than $K$.  To this goal  we will use the generating function method for  the  cycle sum $Z_N$  \cite{Stanley}. We set   $Z_0 = 1$ and for $N\ge 1$
\begin{eqnarray}
\label{A1}
&& Z_N(t_1,\ldots,t_N)  \equiv  \sum_{\sigma}\prod_{k=1}^Nt^{C_k(\sigma)}_k\nonumber\\
&& = \left.  \left(\frac{d }{dx}\right)^N\exp\left\{ \sum_{k=1}^\infty t_k\frac{x^k}{k}\right\}\right|_{x=0},
\end{eqnarray} 
where  $(C_1,\ldots,C_N)$  is the cycle type  of permutation,   with $C_k $ being  the number of cycles of length $k$,  and $t_k$ being  the corresponding control parameter (in the exponent only  the terms up to  $x^N$ contribute). For example,      by setting $t_k =1$ for all $k$  we get $Z_N(1,\ldots,1)=N!$, i.e.,  the number of permutations in $S_N$.  To count all the permutations with the cycle type $(C_1,\ldots,C_K,0,\ldots, 0)$ we must use $t_k=1$ for $1\le k\le K$ and zero otherwise.  The fraction  of such    permutations  can be   estimated for  $N\gg1$  by employing the  following  asymptotic  formula  \cite{S_N_K}
 \begin{eqnarray}
\label{A2}
&& \mathcal{F}^{(K)}_N\equiv \frac{Z_N(1,\ldots,1,0,\ldots,0)}{N!} \nonumber\\
&&  =\frac{1}{N!}\left.\left(\frac{d }{dx}\right)^N\exp\left\{ \sum_{k=1}^K \frac{x^k}{k}\right\}\right|_{x=0}\nonumber\\
 && =   \frac{1+o(1)}{(N!)^\frac{1}{K}(2\pi N)^{\frac{K-1}{2K}}}\frac{\exp\left\{R_{N,K}\right\}}{\sqrt{K}},
\end{eqnarray} 
  with (for $K\ge 2$)
 \begin{eqnarray}
 \label{A3}
 R_{N,K}& =& \sum_{s=1}^{K-1}\frac{\left(\frac{s}{K} +1\right)\ldots \left(\frac{s}{K} +s-1\right)}{s!(K-s)!}N^{\frac{K-s}{K}} \nonumber\\
 &-& \frac{1}{K}\sum_{s=2}^K\frac{1}{s}.
 \end{eqnarray}
 Let us estimate $R_{N,K}$ from above and from below.  The first sum in Eq. (\ref{A3}) dominates (since $K = O(1)$ as $N$ scales up). For  $1\le s\le K-1$ one can see that 
 $(\frac{s}{K} +1)\ldots (\frac{s}{K} +s-1)<s!$, hence 
  \begin{eqnarray}
  \label{A4}
 &&R_{N,K}< \sum_{s=1}^{K-1}\frac{\left(\frac{s}{K} +1\right)\ldots \left(\frac{s}{K} +s-1\right)}{s!(K-s)!}N^{\frac{K-s}{K}}\nonumber\\
 && < \sum_{s=1}^{K-1}\frac{N^{\frac{K-s}{K}}}{(K-s)!} < (e-1)N^{\frac{K-1}{K}}.
 \end{eqnarray}
  
Using the   Stirling approximation \cite{Robbins} 
\[
N!= \sqrt{2\pi N}\left(\frac{N}{e}\right)^N e^{r_N},  \quad r_N > \frac{1}{12N+1},
\]
we obtain  for the fraction of permutations with cycles of length not exceeding $K$ 
 \begin{eqnarray}
 \label{A5}
   \!\!\! \mathcal{F}^{(K)}_N< \frac{1+o(1)}{\sqrt{2\pi N}}\exp\left\{ -N\left[\frac{\ln N - 1}{K} -  \frac{e-1}{N^\frac{1}{K}}\right]  \right\},\nonumber\\
 \end{eqnarray}
For  $K = O(1)$, as $N$ scales up   the  fraction of permutations with the disjoint cycles of lengths not exceeding $K$ is    vanishing   exponentially in $N$.

Let us also work out explicitly   a simple example. Consider the subset of permutations having at least $N-K$ fixed points (i.e., $C_1 \ge  N-K$). Since  the maximal length of a cycle  is equal to  $K$, this example gives the  subset of    permutations having no cycles of length exceeding $K$ (precisely only such  permutations are taken into account by the efficient classical approximations in Refs. \cite{R1,RSP,VS2019}). The total number of such permutations  can be decomposed into disjoint subsets  with $N-K+s$ fixed points with $0\le s\le K$:
\be
D_K(N) = \sum_{s=0}^K \binom{N}{s}d_s,
\en{D_K}
 where  $d_s$ is the number of  permutations of $s$ objects  without   fixed points, i.e., the derangements (see Ref. \cite{Stanley}), given as follows
\begin{eqnarray}
\label{A6}
   d_s = s! \sum_{j=0}^s\frac{(-1)^j}{j!}.
   \end{eqnarray}
 Using that   \mbox{$N!/(N-s)!  =  N^s\left(1 + O(\frac{s^2}{N}) \right)$,}  for $K = O(1)$   we get  the following  estimate on the   total number of  permutations  with at least $N-K$ fixed points   
\be
D_K(N) =  O\left(N^K\right).
\en{A7}
As predicted by Eq. (\ref{A5}) the   fraction   $D_K(N)/N!$ is  exponentially vanishing with $N$.

In conclusion, the fraction of permutations in $S_N$ that have no disjoint  cycle of length greater than $K = O(1)$ is exponentially vanishing as $N$ scales up. 


  \section{ The fraction of weighted permutations   with only lower-order cycles     }
  \label{appAB}

Let us first make an observation on the physical significance of the relative permutation ($\pi$) in      the output probability formula in Eq. (\ref{B1}).  As distinct from the relative permutation  $\pi$,   the other permutation ($\sigma$) gives a  spurious order of identical bosons and   appears also in the classical limit  $\xi \to 0$ of  completely distinguishable bosons (or classical particles, classically  indistinguishable).  For distinguishable bosons (classical particles)    $C_1(\pi) = N$ ($\pi$ is trivial permutation)  and the probability becomes proportional to the matrix permanent of a positive matrix  
 \be
 p_\m(0) = \frac{1}{\m!}\sum_{\sigma}  \prod_{i=1}^N|U_{\sigma(i),l_i}|^2. 
 \en{B2}
 Hence, \textit{only one} (of the two) permutations in the expression for the output probability in Eq. (\ref{B1}) is   mapped to the  the multi-boson interferences, the other one permutes classical particles (see the full theory  in Ref. \cite{Ninter}). This simple fact   also applies  to    the probability in the ideal case of completely indistinguishable bosons 
 \be
 p(1) = \frac{1}{\m!}\left|\sum_{\sigma}  \prod_{i=1}^NU_{\sigma(i),l_i}\right|^2,
 \en{B3}
where, due to bosons being identical particles (no labels),  from the two permutations (in the amplitude and the complex conjugate amplitude) only  one could  correspond to  different multi-boson interferences. The other, quite similarly,  can be accounted for   by the classical particles (classically indistinguishable).

Cycles of length $\ell$ in the  permutation $\pi$ are weighted by $\xi^\ell$ in Eq. (\ref{B1}), since each such cycle involves exactly $\ell$ overlaps of different internal states, e.g., cycle  $k_1\to k_2\to \ldots \to k_{\ell}\to k_1$  corresponds to the product of the overlaps 
\be
\prod_{i=1}^\ell \langle\psi_{k_{i+1}}|\psi_{k_i}\rangle = \xi^\ell, \quad (\ell+1\to  1).
\en{B4}
For example,  a cutoff in the summation  in    Eq. (\ref{B1}) at some maximum order  $\xi^K$ (as in Refs. \cite{R1,RSP,VS2019})   is  equivalent to  retaining only  the subset of permutations $\pi$ with at least $C_1=N-K$ fixed points.  There are   $D_K(N) = O(N^K)$ of such  permutations $\pi$ for $K= O(1)$,  see Eq. (\ref{A7}) of appendix \ref{appA}. This is also a   subset of permutations $\pi$ having no cycles of length greater than $K$.   

Consider  the group of  permutations $\pi\in S_N$,  where each permutation is weighted by $\xi^{N-C_1(\pi)}$, i.e., each cycle is weighted as in Eq. (\ref{B4}) by the two-boson overlap $\xi$.   If we retain only the permutations  with cycles of lengths $\ell\le K$, the discarded   permutations exponentially dominate by  their  number, as shown in appendix \ref{appA},  but with  individual contributions  weighted down by higher powers of the overlap $\xi$.      Our goal is to estimate  the relative fraction of weighted  permutations $\pi$ having no cycles of length greater than $K$, similar  as in Appendix \ref{appA} for the unweighted permutations.  

We start with estimating the total sum of the  weighted permutations in $S_N$. This can be done using the same generating function method, used in Appendix \ref{appA}. Using  the expression for $Z_N$ of  Eq. (\ref{A1}) with $t_1=1/\xi$ and $t_{k}=1$ for all $k\ge 2$,  we obtain  
\begin{eqnarray}
\label{B5}
&&\mathcal{Z}_N(\xi) \equiv \sum_{\pi} \xi^{N-C_1(\pi)} = \xi^NZ_N(1/\xi,1,\ldots,1)\nonumber\\
&&= \xi^N\left.  \left(\frac{d }{dx}\right)^N \frac{e^{(\frac{1}{\xi}-1)x}}{1-x}\right|_{x=0}= \xi^N N! \sum_{n=0}^N\frac{(1/\xi-1)^n}{n!}, \nonumber\\
\end{eqnarray}
 where we have used Leibniz's rule for the $N$-order derivative of a product.  
  
Let us now estimate the contribution $\mathcal{Z}^{(K)}_N(\xi)$ to $\mathcal{Z}_N(\xi)$ of Eq. (\ref{B5}) coming from the permutations $\pi$ with the disjoint cycles of lengths not exceeding $K$.  The simplest bound follows from setting   $\xi=1$   (completely indistinguishable bosons):
\be
\mathcal{Z}^{(K)}_N (\xi) \le    \mathcal{Z}^{(K)}_N (1),
 \en{B9}
 where $  \mathcal{Z}^{(K)}_N (1)$ is  the total number of permutations with the disjoint cycles of lengths $\ell \le K$,  considered in appendix \ref{appA}. 
Hence,  utilizing the asymptotic bound on $\mathcal{F}^{(K)}_N$ from Eqs. (\ref{A5}) and (\ref{B9}) we obtain (observing that  $1/\xi-1>0$)
\begin{eqnarray}
\label{B10}
&& \frac{\mathcal{Z}^{(K)}_N(\xi)}{\mathcal{Z}_N(\xi)} <  \frac{\mathcal{F}^{(K)}_N}{\xi^{N}}   \\
 && < \frac{1+o(1)}{\sqrt{2\pi N}}\exp\left\{ -N\left[\frac{\ln N - 1}{K} -  \frac{e-1}{N^\frac{1}{K}}-\ln \xi^{-1}\right]  \right\}.\nonumber
 \end{eqnarray}
 Therefore,  as  in the  noiseless case   ($\xi = 1$),   the fraction    of permutations with the disjoint cycles of lengths not exceeding $K$ is exponentially vanishing in $N$ for $K=O(1)$ and any {constant noise} $\xi^{-1} = O(1)$. 
 
  Thus,   for constant noise (parameter $\xi$ bounded from below) the  permutations with the lower-order disjoint  cycles, i.e., with lengths bounded by  $K = O(1)$,  even if  unweighted by the overlap $\xi$,   correspond to an exponentially vanishing  fraction of all  weighted  permutations.   
 
  \section{Output probabilities  with  contribution from the higher-order interferences}
  \label{appC}
\label{appC}

Here we look for the output probabilities which reveal the  contribution from higher-order cycles (higher multi-boson interferences). To this goal, consider a probability $P_\Omega$ that all the bosons are detected in some subset $\Omega$ of $M$  output ports. Such probability is obtained by summation  in Eq. (\ref{B1}) over the output configurations $\m$, i.e., the occupations  of output ports $l_i\in \Omega$, $i=1,\ldots, N$, or equivalently, by summation of $\frac{\m!}{N!}p_\m$ over    \textit{independent} output port  indices $l_i\in\Omega$, $i=1,\ldots, N$.  Let us introduce a  positive semi-definite Hermitian matrix 
\be
A_{kj}\equiv \sum_{l\in \Omega} U_{kl}U^*_{jl}. 
\en{C1}
Then, performing the summation as above indicated, we get  the probability $P_\Omega$  as follows 
\begin{eqnarray}
\label{C2}
&& P_\Omega =  \frac{1}{N!}\sum_{\sigma} \sum_\pi \xi^{N-C_1(\pi)} \prod_{i=1}^NA_{\sigma(i),\pi\sigma(i)}\nonumber\\
 && = \sum_\pi \xi^{N-C_1(\pi)} \prod_{i=1}^NA_{i,\pi(i)},
\end{eqnarray} 
where now there is only one relative   permutation $\pi$, since $\sigma$ has no effect (as seen by reordering the terms in the product; this is a consequence of the different physical meaning of the two permutations, discussed in appendix \ref{appAB}).  We can simplify the result even more, by observing that the $N-C_1(\pi)$ in the $\xi$-factor counts the number of off-diagonal elements (i.e., $\pi(i)\ne i$) in the matrix $A$. Hence, by introducing the rescaled matrix as follows
\be
\mathcal{A}_{kj} = \left\{\begin{array}{cc} A_{kk},  & j=k\\
\xi A_{kj}, & j\ne k \end{array} \right.,
\en{C3}
we obtain the probability in the form of a matrix permanent 
\be
P_\Omega = \mathrm{per}\mathcal{A}. 
\en{C4}
Such probabilities can be  approximated by polynomial  classical computations  \cite{PermHerm}.



\begin{thebibliography}{99}

 

\bibitem{AA} S. Aaronson and A. Arkhipov,  The computational complexity of linear optics.  \href{https://doi.org/10.4086/toc.2013.v009a004}{\textit{Theory of Computing} \textbf{9},  143 (2013).}

\bibitem{Cliffords} P. Clifford, and R. Clifford. The Classical Complexity of Boson Sampling. 
\href{https://doi.org/10.1137/1.9781611975031.10}{\textit{Proceedings of the 2018 Annual ACM-SIAM Symposium on Discrete Algorithms} pp. 146–55.} 

\bibitem{20ph60mod} H.  Wang, J.  Qin, X. Ding, M.-C. Chen, S. Chen, X. You, Y.-M. He, X. Jiang, L. You, Z. Wang, C. Schneider, J. J. Renema, S.~Höfling, C.-Y. Lu, and J.-W. Pan.
 Boson Sampling with 20 Input Photons and a 60-Mode Interferometer in a $10^{14}$-Dimensional Hilbert Space. 
 \href{https://doi.org/10.1103/PhysRevLett.123.250503}{\textit{Phys. Rev. Lett.} \textbf{123},  250503 (2019).} 

\bibitem{ExpGBS2}  H.-S. Zhong \textit{et al},  Quantum computational advantage using photons, Science \textbf{370},   1460 (2020). 


\bibitem{GBS144} H.-S.  Zhong, \textit{et al.} Phase-Programmable Gaussian Boson Sampling Using Stimulated Squeezed Light.  \textit{Phys. Rev. Lett. } \textbf{127}, 180502 
(2021). 






\bibitem{VS14}      V. S. Shchesnovich.  Sufficient condition for the mode mismatch of single photons for scalability of the boson-sampling computer.
\href{https://doi.org/10.1103/PhysRevA.89.022333}{\textit{Phys. Rev.  A}  \textbf{89},   022333  (2014).}
 

\bibitem{Brod} S. Aaronson and D. J. Brod. BosonSampling with lost photons. 
\href{https://doi.org/10.1103/PhysRevA.93.012335}{\textit{Phys. Rev.  A}  \textbf{93}, 012335 (2016).}


\bibitem{LP}  A. Leverrier and R. Garc{\'i}a-Patr{\'o}n. Analysis of circuit imperfections in BosonSampling. 
\href{https://dl.acm.org/doi/abs/10.5555/2871401.2871409}{\textit{Quant. Inf. \& Computation} \textbf{15}, 489-512 (2015). }

\bibitem{KK}  G. Kalai and G. Kindler. Gaussian Noise Sensitivity and BosonSampling. \href{https://arxiv.org/abs/1409.3093}{arXiv:1409.3093 [quant-ph].} 

 \bibitem{Arkh}  A. Arkhipov. BosonSampling is robust against small errors in the network matrix. 
\href{https://doi.org/10.1103/PhysRevA.92.062326}{\textit{Phys. Rev.  A}  \textbf{92}, 062326 (2015).}


\bibitem{Bblog} S. Aaronson. \href{https://scottaaronson.blog/?p=5868}{Gaussian BosonSampling, higher-order correlations, and spoofing: An update, Blog post, October 10, 2021.}


 
\bibitem{K1} S. Rahimi-Keshari,  T.  C. Ralph, and C.  M. Caves. Sufficient Conditions for Efficient Classical Simulation of Quantum Optics.
\href{https://doi.org/10.1103/PhysRevX.6.021039}{\textit{Phys. Rev. X} \textbf{6}, 021039 (2016).} 

\bibitem{R1}   J. J. Renema, A. Menssen, W. R. Clements, G. Triginer, W. S. Kolthammer, and I.~A.~Walmsley. Efficient Classical Algorithm for Boson Sampling with Partially Distinguishable Photons. \href{https://doi.org/10.1103/PhysRevLett.120.220502}{\textit{Phys. Rev. Lett.} \textbf{120}, 220502  (2018).}

\bibitem{OB} M. Oszmaniec and D. J. Brod. Classical simulation of photonic linear optics with lost particles. 
\href{https://doi.org/10.1088/1367-2630/aadfa8}{\textit{New J. Phys.} \textbf{20}, 092002 (2018).}

\bibitem{PRS} R. Garc\'ia-Patr\'on, J. J. Renema, and V.~S.~Shchesnovich. Simulating boson sampling in lossy architectures.
\href{https://doi.org/10.22331/q-2019-08-05-169}{\textit{Quantum} \textbf{3}, 169 (2019).}

\bibitem{NonUnifLoss} 	  D. J. Brod and M. Oszmaniec. Classical simulation of linear optics subject to nonuniform losses.
\href{https://doi.org/10.22331/q-2020-05-14-267}{\textit{Quantum} \textbf{4},  267 (2020).}

\bibitem{RSP}  J. J. Renema, V. S. Shchesnovich, and R. Garc\'ia-Patr\'on. Classical simulability of noisy boson sampling.
\href{https://arxiv.org/abs/1809.01953}{arXiv:1809.01953 [quant-ph].} 
 
\bibitem{VS2019} V. S. Shchesnovich. Noise in boson sampling and the threshold of efficient classical simulatability.
\href{https://doi.org/10.1103/PhysRevA.100.012340}{\textit{Phys. Rev. A} \textbf{100}, 012340  (2019).}

\bibitem{MRRT} A. E. Moylett,  R. Garc\'ia-Patr\'on, J. J. Renema, and P. S. Turner. Classically simulating near-term partially-distinguishable and lossy boson sampling.
\href{https://doi.org/10.1088/2058-9565/ab5555}{\textit{Quantum Sci. Technol.}  \textbf{5},  015001  (2020).}

\bibitem{Ablog} S. Aaronson. \href{https://scottaaronson.blog/?p=5159}{Chinese BosonSampling experiment: the gloves are off.  Blog post, December 16, 2020.}


\bibitem{myQuantum}  V. S. Shchesnovich.  Distinguishing noisy boson sampling from classical simulations. Quantum \textbf{5}, 423 (2021).



\bibitem{Ninter} V. S. Shchesnovich and  M. E. O. Bezerra. Collective phases of identical particles interfering on linear multiports.
\href{https://doi.org/10.1103/PhysRevA.98.033805}{\textit{Phys. Rev. A} \textbf{98}, 033805 (2018).}

 


\bibitem{PermHerm} L. Chakhmakhchyan, N. J. Cerf, and R.~Garc\'ia-Patr\'on.
Quantum-inspired algorithm for estimating the permanent of positive semidefinite matrices. 
\href{https://doi.org/10.1103/PhysRevA.96.022329}{\textit{Phys. Rev.  A} \textbf{96}, 022329 (2017).} 


\bibitem{ClassSimGBS} B.  Villalonga,  M.  Y.  Niu,  L. Li, H. Neven,  J. C. Platt, V. N. Smelyanskiy, and S. Boixo. Efficient approximation of experimental Gaussian boson sampling.  	arXiv:2109.11525 [quant-ph]. 


\bibitem{Stanley}   R. P. Stanley, \textit{Enumerative Combinatorics}, 2nd ed., Vol. 1 (Cambridge University Press, 2011). 


\bibitem{S_N_K}  M. A. Evgrafov. \textit{Asymptotic Estimates and Entire Functions}.  (Dover Publications, 2020). See problem 3 in Paragraph 1.6. 

\bibitem{Robbins} H. Robbins. A Remark on Stirling's Formula. The Amer. Math. Monthly, \textbf{62}, 26 (1955). 

\end{thebibliography}
\end{document}